\documentclass[amsfonts,amssymb,aps,prb]{revtex4-2}

\usepackage[english]{babel}
\usepackage{xcolor}
\usepackage[colorlinks = true,
            linkcolor = blue,
            urlcolor  = blue,
            citecolor = blue,
            anchorcolor = blue]{hyperref}
\usepackage{amsmath,amssymb}
\usepackage{geometry,graphicx,cleveref}
\usepackage{tikz}
\usepackage{subcaption}

\newcommand{\mathd}{\mathrm{d}}
\newcommand{\mathe}{\mathrm{e}}

\newcommand{\tmop}[1]{\ensuremath{\operatorname{#1}}}

\begin{document}

\title{Exciton crystal melting and destruction by disorder in bilayer quantum hall system
with total filling factor one}

\author{Zhengfei Hu and Kun Yang}

\affiliation{Department of Physics and National High Magnetic Field Laboratory,
  Florida State University, Tallahassee, Florida 32306}

\date{\today}

\begin{abstract}
  Bilayer quantum hall system with total filling factor 1 was studied in the
  regime of heavy layer imbalance in a recent transport experiment [Zeng2023,
  arXiv:2306.16995], with intriguing new findings. We demonstrate in this paper
  that 1) the exciton Wigner crystal in this regime can melt into a superfluid
  phase, giving rise to re-entrant superfluid behavior; 2) in the presence of
  disorder, electron and hole Wigner crystals in the two layers go through a
  locking/decoupling transition as layer separation increases, resulting in a
  sudden change in the counter flow conductance. Comparison will be made with
  the findings of experiments.
\end{abstract}

{\maketitle}

\section{Introduction}

Bilayer quantum hall system with total filling factor $\nu_1 + \nu_2 = 1$ has been
actively studied for several decades
\cite{Wen1992,EZAWA1992,Eisenstein_1992,Wen1993,Yang1994,Butov_1994,
  Moon1995,Yang1996,Lilly_1998,yang2001,Butov_2002,Eisenstein2004,Kellogg_2004,
  Tutuc_2004,Wiersma_2004,Seradjeh_2008,Tiemann_2008,Weitz_2010,Barlas_2012,
  Nandi_2012,Eisenstein_2014,Liu_2017,Li_2017,Burg_2018}. The long-lasting
interest in it is due to its extremely rich phase diagram and the fascinating
physics associated with the novel phases and transitions among them, which is
yet to be exhausted.  A recent transport experiment {\cite{zeng2023}} focused on
a regime that is under-explored before, namely when the two layers are heavily
imbalanced, such that $\Delta\nu=\nu_1 - \nu_2 \lesssim 1$, namely $\nu_2\ll 1$ is the minority layer
of electrons, and the hole filling factor in the majority layer 1 is
$1-\nu_1 = \nu_2$. The experiment observed an exciton superfluid-insulator
transition predicted more than 20 years ago {\cite{yang2001}}, and revealed some
new surprises. The purpose of this work is to provide theoretical understandings
of two of the new findings.

We start by briefly summarising the relevant observations and basic
idea/conclusion of our theoretical work. The experimentalists pass a (drive)
current through one of the layers, and measure the current and/or voltage
response of the same as well as opposite layer; the latter corresponds to drag
response \cite{Gramila_1991}. Symmetric and antisymmetric combinations of these
responses form normal and counter flow response functions; the latter is usually
attributed to the flow of interlayer excitons which are bound pairs of electron
in one layer and hole in the other, assuming they are present and dominate the
counter flow transport channel. Bounding between electrons and hole results in
the suppression of free charge carrier, and hence an insulating state of net
in-plane charge transport. The excitons, on the other hand, may either condense
to form a superfluid (SF), or crystallize and form an insulating Wigner crystal
(WC) state. We will demonstrate that under appropriate conditions an exciton
Wigner crystal may melt into a superfluid state, giving rise to re-entrant
superfluid behavior in the counter flow channel seen in the experiment. We
further demonstrate that presence of uncorrelated disorder potential in the two
layers can disrupt the formation of the interlayer excitons, driving a
transition between exciton Wigner crystal and decoupled electron and hole Wigner
crystals in each layer. This transition manifests itself in some transport
anomalies observed in the counterflow channel. It should be noted that, there
could be other phase transitions, e.g., transition between decoupled fractional
quantum hall phase and superfluid at $\Delta\nu = 1/3$ {\cite{Chen_2012}}. They will
compete with the Wigner crystal phase when $|\Delta\nu|$ goes away from 1.

The rest of the paper is organized as following.  In Sec. \ref{part1}, we
calculate the critical temperature of bilayer exciton superfluid using two
previously established effective models {\cite{Moon1995,yang2001}} at layer
imbalance $1 - | \Delta \nu | \ll 1$, and demonstrate it is often higher than the melting
temperature of exciton Wigner crystal. As a result the crystal melts into a
superfluid when this is the case. In Sec. \ref{part2} we consider the interplay
of disorder and interlayer coupling and analyze the competition between
them. Clearly interlayer Coulomb coupling drives formation of interlayer
excitons, while uncorrelated disorder favors formation of decoupled electron and
hole Wigner crystals in each layer. By comparing the energy gains from exciton
formation and uncorrelated electron and hole WC distortion in the two layers, we
obtain the phase diagram of the system. Some concluding remarks are provided in
Sec. \ref{summary}.

Unless otherwise stated, magnetic length is assumed to be the length scale, i.e. $l_B=1$.

\section{Exciton superfluid and melting of Wigner crystal \label{part1}}

We start by discussing the phases relevant to this section. It is
well-established that single layer 2-dimensional electron gas forms a Wigner
crystal at zero temperature for small $\nu$
\cite{Wigner_1934,Maki_1983,Yoshioka_1983,Lam_1984,MacDonald_1987,
  Goldman_1990,Senatore_1994,Manoharan_1994,Shayegan_1996,Kamilla_1997,
  Yang_2001wc,Ye_2002,Chen_2006,Monceau_2012,Deng_2019,Smole_ski_2021,Tsui_2024}.
Putting two layers together and holding the total filling factor
$\nu_1 + \nu_2 = 1$, the electron (in the minority layer 2) and hole (on the
majority layer 1) Wigner crystals with identical structure lock into an exciton
crystal {\cite{yang2001}}, which may melt due to either quantum or thermal
fluctuations. Comparisons between drag current versus drive current, and
parallel flow versus counter flow conductance, indicate that the resulting zero
temperature phase is indeed correlated between the two layers
\cite{zeng2023}. Electrons in one layer and holes in the other tend to bind and
condense into an exciton superfluid when $d$ is small and $1-|\Delta \nu|$ is not too
close to 1, and form an exciton Wigner crystal otherwise; see orange line of
\Cref{fig:2} for schematic zero temperature phase diagram near $\Delta \nu = 1$. With
increasing temperature the exciton Wigner crystal melts into a liquid. We find,
surprisingly, that under appropriate conditions the resultant liquid state may
be a superfluid.

To understand this we go back to zero temperature, where the exciton superfluid
and Wigner crystal phases compete with each other. They are (most likely)
separated by a 1st order phase boundary, allowing us to consider thermal effects
on them at finite temperature separately. As discussed earlier the exciton
Wigner crystal melts into a liquid at some melting temperature which we estimate
below. The exciton superfluid, on the other hand, goes through a
Kosterlitz-Thouless (KT) transition and becomes a normal fluid. If the
superfluid critical (KT) temperature is lower than the melting temperature, we
expect WC melts into a normal fluid which is the usual situation. If it turns
out the KT temperature is higher than the melting temperature, we conclude that
the WC melts into a superfluid instead, resulting in re-entrant
superfluidity. The resultant (schematic) phase diagram takes the form of
\Cref{fig:1}. Our results compare favorably with those of {\cite{zeng2023}}.

To determine the phase diagram we start by calculating the superfluid stiffness
which determines the KT temperature of the superfluid phase, and then compare it
with the melting temperature of the WC.




\subsection{Phase stiffness and Kosterlitz-Thouless temperature of exciton superfluid}

When $\Delta \nu$ is fixed, the low temperature superfluid behavior can be described by
an effective XY model. In this section we calculate the phase stiffness from two
different models: spin $1 / 2$ easy-plane ferromagnet {\cite{Moon1995}} and
dilute exciton {\cite{yang2001}}. Once the phase stiffness $\rho_s$ is obtained,
critical temperature of SF is bounded by $T_c = \frac{\pi \rho_s}{2}$. It turns
out in the vicinity of $\Delta \nu = 1$, two models lead to the same result. Let 
$Q^2 = e^2 / (4 \pi \epsilon)$ for simplicity.

\subsubsection{spin-1/2 easy-plane ferromagnet}

To begin with, we setup the notations here. Let
$\nu_1 = \nu_{\uparrow} = 1 - \delta, \nu_2 = \nu_{\downarrow} = \delta$, we
have
$\Delta \nu = (1 - 2 \delta) = \cos \theta = 2 (S_{\uparrow} - S_{\downarrow}) =
m^z$ and $\delta = \frac{1 - \Delta \nu}{2} = \sin^2 (\theta / 2)$, density of
electron in one layer
\begin{equation}
\label{eq:e-density}
n = \delta / 2 \pi = \sin^2 (\theta / 2) / 2 \pi
\end{equation}

The gradient energy density of xy components of local spin is
\begin{equation}
  \frac{\rho_E}{2}  [(\nabla m^x)^2 + (\nabla m^y)^2],
\end{equation}
where
$\rho_E = - \frac{\nu}{32 \pi^2} \int_0^{\infty} V_k^E h (k) k^3 \mathd k$, and
$V_k^E = V_k^A e^{- kd}, V_k^A = \frac{2 \pi Q^2}{k}$ are fourier transforms of
intralayer Coulomb potential and interlayer Coulomb potential respectively
{\cite{Moon1995}} .
$h (k) = \frac{\nu}{2 \pi} \int \mathd^2 r (g (r) - 1) \exp (- i \mathbf{k} \cdot
\mathbf{r})$ and $g (r) = \langle c^{\dag} (\mathbf{r}) c (0) \rangle$ are particle-hole
correlation of Laughlin function in momentum space and real space.

For $\nu = 1$, $g (r) = \exp (- r^2)$ and $h (k) = - \exp (-
\frac{|k|^2}{2})$, hence we have
\begin{equation}
  \rho_E = - \frac{Q^2}{16 \pi }  \left[ d - \sqrt{\frac{\pi}{2}}  (d^2 + 1)
  e^{\frac{d^2}{2}} \tmop{erfc} \left( d / \sqrt{2} \right) \right] \equiv
  \frac{Q^2 f (d)}{16 \pi},
\end{equation}
where $d$ is the interlayer spacing, 
$f (d) = \sqrt{\frac{\pi}{2}}  (d^2 + 1) e^{\frac{d^2}{2}} \text{erfc}
\left( d / \sqrt{2} \right) - d$ and $\tmop{erfc} (x) = 1 - \tmop{erf} (x)$ is
the complementary error function.

After we obtain $\rho_E$, phase stiffness of XY spin is $\rho_s = \rho_E
\sin^2 \theta$
\begin{equation}
  \rho_s^{\tmop{XY}} = \frac{Q^2 f (d)}{4 \pi}  \frac{\sin^2 (\theta)}{4} =
  \frac{Q^2 f (d)}{8 \pi}  \frac{1 - (\Delta \nu)^2}{2}, \label{stiffness-xy}
\end{equation}
and the critical temperature $T_{\mathrm{KT}} \lesssim \frac{\pi}{2} \rho_s$.

\subsubsection{Dilute dipolar exciton}

From {\cite{yang2001}} inverse effective mass of exciton is
\begin{equation}
  m (d)^{- 1} = \frac{Q^2}{2}  \int_0^{\infty} x^2 \mathe^{- xd - x^2 / 2}
  \mathd x = \frac{Q^2}{2}  \left( \sqrt{\frac{\pi}{2}}  (d^2 + 1)
  e^{\frac{d^2}{2}} \text{erfc} \left( d / \sqrt{2} \right) - d \right) =
  \frac{Q^2}{2} f (d)
\end{equation}
Boson spectrum given by Bogoliubov theory (see e.g. chap18 of
{\cite{girvin2019}}) is
\begin{equation}
  E_{\mathbf{k}} = \sqrt{\epsilon_{\mathbf{k}}^2 + 2 n \tilde{V}_{q = 0}
  \epsilon_{\mathbf{k}}} \xrightarrow{k \rightarrow 0} \sqrt{2 n \tilde{V}_0
  \epsilon_{\mathbf{k}}} = \hbar v_s k,
\end{equation}
where the effective interaction $\tilde{V}_k = 2 \Delta V_k - \frac{2}{N}
\sum_{\mathbf{q}} \Delta V_q e^{- q^2 / 2}, \Delta V = V^A - V^E$
{\cite{yang2001}}. The Goldstone mode velocity $v_s = \sqrt{\frac{n
\tilde{V}_0}{m}}$ is also reported in {\cite{yang2001}}.

Thereafter superfluid phase stiffness $\rho_s = \frac{n}{m}$ can be obtained from
$n\mathbf{v}_s = \rho_s \nabla \theta$ and $\mathbf{v}_s = \nabla \theta / m$ where
$n$ is given in (\ref{eq:e-density}).
\begin{equation}
  \rho_s^{\tmop{exciton}} = \frac{Q^2 f (d)}{4 \pi} \sin^2  \frac{\theta}{2} =
  \frac{Q^2 f (d)}{8 \pi}  (1 - \Delta \nu), \label{eq:stiffness-exciton}
\end{equation}
This expression of superfluid density coincides with the result
\eqref{stiffness-xy} when $\Delta \nu \rightarrow 1$ (or $\theta \rightarrow 0$) since
$\frac{1 - (\Delta \nu)^2}{2} = 1-\Delta \nu -(1-\Delta \nu)^2 / 2 \simeq 1-\Delta \nu$.

We will stick to \Cref{eq:stiffness-exciton} and use
$T_{\mathrm{KT}}=\pi \rho_s^{\tmop{exciton}} / 2$ as our estimate of KT temperature
\begin{equation}
  t_{\mathrm{KT}} \equiv T_{\mathrm{KT}}/Q^2 = \frac{f (d)}{16} (1 - \Delta \nu), \label{eq:t-c}
\end{equation}

\subsection{Melting temperature of exciton Wigner crystal and phase diagrams}

In this subsection we compare melting temperature of exciton Wigner crystal,
$T_{m}$, with the KT temperature estimated above, and determine the finite
temperature phase diagram of the system.

\begin{figure}[!h]
    \begin{subfigure}[c]{0.48\textwidth}
        \resizebox{\linewidth}{!}{
            \includegraphics{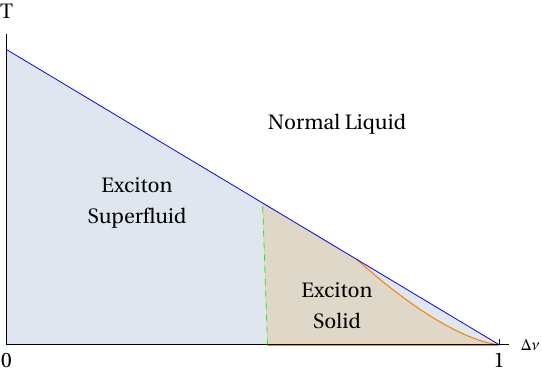}
        }
        \caption{}
        \label{fig:1-a}
    \end{subfigure}
    \begin{subfigure}[c]{0.48\textwidth}
        \resizebox{\linewidth}{!}{
            \includegraphics{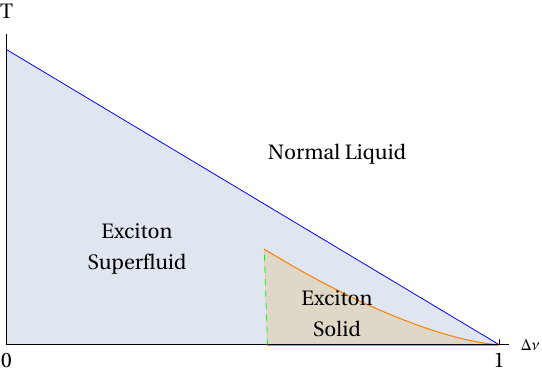}
        }
        \caption{}
        \label{fig:1-b}
    \end{subfigure}
    \caption{Finite temperature phase diagrams near $\Delta \nu = 1$ based on
      \Cref{eq:t-c,eq:t-m-exciton}. The green dashed line is the natural
      extension of the zero temperature phase boundary between exciton
      superfluid and Wigner crystal phases. Blue line is the superfluid KT
      temperature.  Orange line is the melting curve of exciton Wigner crystal.
      (a) Case with $d > d_c \simeq 2$ in which the exciton Wigner crystal can melt
      into either superfluid or normal liquid, depending on $\Delta \nu$. (b) Case with
      $d < d_c$ where the exciton Wigner crystal can only melt into a
      superfluid. It should be noted that region far from $\Delta\nu = 1$ shall not be
      taken too literally.\label{fig:1}}
\end{figure}

\begin{figure}[!h]
            \includegraphics{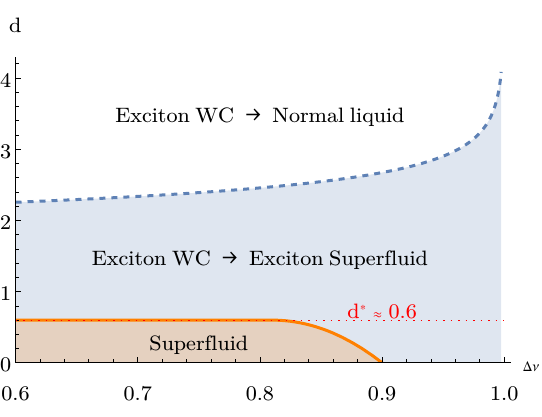}  
    \caption{Schematic zero temperature phase diagrams near $\Delta \nu = 1$. 
    Orange region denotes the superfluid phase
      which, due to disorder, terminates before $\Delta \nu = 1$ is reached.
      Orange solid line is the schematic zero temperature phase boundary
      between superfluid and Wigner crystal. Blue and blank region are both Wigner
      crystal at zero temperature, while the blue one melts into superfluid with
      increasing temperature and the blank one melts into normal liquid (see the
      arrows in the right panel). The blue dashed line is obtained by equating 
      \Cref{eq:t-m-exciton} with \Cref{eq:t-c}. The red dotted line marked by $d^{\ast} = 0.6$,
      reported in {\cite{yang2001}}, is obtained by comparing correlation energy
      per exciton in superfluid phase and kinetic energy in crystal phase, above which superfluid
      phase is unfavored. \label{fig:2}}  
\end{figure}

The melting temperature of classical exciton Wigner crystal was reported to be
$T_{m} \approx 0.0907 \frac{d^2 Q^2}{a^3}$ \cite{von_Gr_nberg_2004,Kalia_1981}.
Relation $a = [\frac{\sqrt{3}}{8\pi} (1-\Delta \nu)]^{-1/2}$ can obtained from
$\frac{1-\Delta \nu}{2} = \frac{n_e}{1/2\pi}$ where $n_e = 2/(\sqrt{3} a^2)$.  We then
have dimensionless temperatures
\begin{equation}
    t_{m} = 0.0907 d^2 [\frac{\sqrt{3}}{8\pi} (1-\Delta \nu)]^{3/2} 
    \label{eq:t-m-exciton}
\end{equation}
where $t_m=T_m / Q^2$. Compare \Cref{eq:t-m-exciton} with \Cref{eq:t-c}, we are
able to determine the finite temperature phase diagrams \Cref{fig:1} for two
different situations, both of which are included in the zero temperature phase
diagram \Cref{fig:2}. Two situations are separated by $d_c \simeq 2$. When
$d>d_c$ the Wigner crystal could melt into either superfluid or normal liquid,
otherwise it only melts into superfluid.  In the dilute limit
$1-\Delta \nu \ll 1$, the exciton Wigner crystal always melts into a superfluid phase
since $t_{m} < t_{\mathrm{KT}}$ is always true.

Treating WC as classical leads to an overestimation of $T_m$, because quantum
fluctuation tends to lower $T_m$ as well. Since our goal is to demonstrate the
possibility of $T_m < T_{\mathrm{KT}}$, they are justified, and does not change
the phase diagram qualitatively. A more serious issue is neglecting the effects
of disorder, which are very important when $\Delta\nu\rightarrow 1$, where the excitons are
destroyed. This is the focus of the next section. The resultant phase there is a
single-layer integer quantum Hall state, which dominates the experimental phase
diagram there. One should keep this in mind when comparing with the theoretical
phase diagrams in this section obtained {\em without} taking these into account.

\section{Locking-decoupling transition of bilayer Wigner crystal\label{part2}}

In the previous section we discussed various phases interlayer excitons can
form, and neglected the effects of disorder. Ref. \cite{zeng2023} finds a single
layer integer quantum Hall state when $\Delta\nu$ is very close to 1, in which the two
layers are essentially decoupled. They also report evidence of a transition into
the exciton Wigner crystal phase discussed above. We argue below the existence
of the decoupled phase is stabilized by disorder, which also drives the
transition. In the absence of disorder potential, the electron and hole WCs in
the two layers always align themselves with each other to minimize the Coulomb
energy, resulting in the exciton WC \cite{yang2001}. On the other hand, disorder
potential, which is different in the two layers (assumed to be uncorrelated for
simplicity), distorts the two WCs in uncorrelated ways, which tends to disrupt
the formation of excitons and decouple the two layers. By comparing the energy
gain and loss between disorder potential energy and interlayer Coulomb energy,
we are able to obtain the transition line for the two layers to become
locked/decoupled.

\subsection{Disorder potential energy}

We introduce the Gaussian white noise random potential $V_i(\mathbf{r})$ that is
uncorrelated between the two layers:
\begin{equation}
    \langle V_i(\mathbf{r}) V_j(\mathbf{r'}) \rangle =
    \Delta^2 \delta (\mathbf{r-r'}) \delta_{i j},
    \label{eq:random-potential}
\end{equation}
where $i,j=1,2$ are layer indices. Pinning length $R$ of 2d Wigner crystal,
defined as $\langle [u(0)-u(R)]^2 \rangle \simeq a^2 $ where $a$ is the lattice constant and
$u(R)$ is the field of lattice distortion, is given by balancing the energy gain
of random potential and energy cost of lattice distortion
{\cite{Chitra2001,Fogler2000}}
\begin{equation}
  Rn_e \Delta = ca^2, \label{balance-random-shear}
\end{equation}
where $n_e = 1 / (A_c a^2), A_c=\sqrt{3} / 2$ is the density of electron, $c$ is
the shear modulus.  Left and right hand sides of this equation stand
respectively for random potential energy gain and elastic energy cost due to
lattice distortion. Since this amount of energy is for a region of linear size
$R$, dividing by $R^2$ we obtain the density of random potential energy (for
convenience in density comparison we keep one factor of $n_e$ here)
\begin{equation}
  \varepsilon_r = \frac{Rn_e \Delta}{R^2} = \frac{n_e \Delta^2}{c A_c a^4}.
\end{equation}
For single layer Wigner crystal of electron-type interaction and dipole-type interaction
we simply take the shear modulus from {\cite{Bruun2014}}
\begin{equation}
  \begin{array}{ll}
    c_1 (d \lesssim a) \approx 2.5 \frac{D^2}{a^5} & \tmop{dipole}\\
    c_2 = 0.3 \frac{Q^2}{a^3} & \tmop{charge}
  \end{array}
\end{equation}
where $Q^2 = \frac{e^2}{4 \pi \epsilon}, D^2  =
\frac{e^2 d^2}{4 \pi \epsilon}$. Transition from coupled to decoupled picture lowers the disorder potential energy (density) by 
\begin{equation}
  \Delta \varepsilon_{r} = \frac{2 n_e \Delta^2}{c_1 A_c a^4} - \frac{n_e 
  \left( \sqrt{2} \Delta \right)^2}{c_2 A_c a^4} = \frac{2 n_e \Delta^2}{A_c a^4} 
  \left( \frac{1}{c_1} - \frac{1}{c_2} \right),
\end{equation}
where $\sqrt{2} \Delta$ is the effective random potential strength seen by the
bilayer (since $V(\mathbf{r})=V_1(\mathbf{r})+V_2(\mathbf{r})$ has $\langle
V(\mathbf{r}) V(\mathbf{r'}) \rangle = 2 \Delta^2 \delta(\mathbf{r} - \mathbf{r'})$). On the
other hand, in $d \rightarrow \infty$ the interlayer Coulomb energy is diminished
and what we have is merely two copies of single layer Wigner crystal. Therefore 
$c_1(\infty) = 2 c_2$. In this limit $\Delta E_{r}$ is exactly half of
that for individual pinning. In practice for a specific $d$ in experiments, the
effective spacing $d/a$ has an upper bound $d/\sqrt{2}$, which is generally
smaller than 1 (see below). For such considerations, we will simply take the
dipole approximation $c_1=2.5 D^2 / a^5$.

\begin{equation}
    \Delta \varepsilon_{r} = \frac{2 n_e \Delta^2}{A_c Q^2 a} 
  \left( \frac{1}{0.3} - \frac{1}{2.5 d^2/a^2} \right)
  = q \frac{n_e Q^2}{a} \left( \frac{1}{0.3} - \frac{1}{2.5 d^2/a^2} \right), \label{eq:e-r}
\end{equation}

where 
\begin{equation}
    q = \frac{2 \Delta^2}{A_c Q^4} = \frac{4 \Delta^2}{\sqrt{3} Q^4} \label{eq:def:q}
\end{equation}
is the dimensionless random potential strength.

\subsection{Interlayer correlation energy cost}

As we demonstrated above, the system can lower the disorder potential energy by distorting the electron and hole WCs in the two layers {\em independently}, compared to that of the exciton WC. Doing that, however, decouples the two layers and destroy the excitons, resulting in an increase in the interlayer Coulomb interaction energy. In this subsection we calculate this energy cost.

In this subsection we let $\frac{Q^2}{a}$ be energy scale and $a$, the lattice
constant of 2d triangular lattice, be length scale. We are evaluating the interlayer
correlation energy difference of Wigner crystal vs. homogeneous electron gas (since random
relative distribution of charges in one layer is seen on average as homogeneous gas of
charge by the other layer), i.e.
\begin{equation}
  \Delta E_e = \int \mathd \mathbf{r} [g_1 (\mathbf{r}) - g_2 (\mathbf{r})] 
  \frac{-1}{\sqrt{r^2 + d^2}} = \int \mathd \mathbf{r} \left[ \sum_i \delta
  (\mathbf{r}-\mathbf{R}_i) - 1 / A_c \right]  \frac{1}{\sqrt{r^2 + d^2}},
\end{equation}
where $g_1 (\mathbf{r}) = 1 / A_c, g_2 (\mathbf{r}) = \sum_i \delta
(\mathbf{r}-\mathbf{R}_i)$, $A_c = \sqrt{3} /
2$ is the area of unit cell. Compared with \Cref{eq:e-r}, a transition between
locked/decoupled phase will be determined.

In the small $d$ limit, apart from a divergent $1 / d$ term, this energy
difference is the classic problem of static energy of 2d Wigner crystal.
That is (see e.g. \cite{Bonsall1977,Borwein1988})
\begin{equation}
  \lim_{d \rightarrow 0} [\Delta E_e (d) - 1/d] = - 4.213423 ,
\end{equation}
We now calculate this energy difference for general $d$. 
Let $\Delta E_e = E_0 + E_1 + E_2$, where $E_0 = 1 / d$ and
\begin{equation}
  \begin{array}{ccl}
    E_1 & = & \frac{1}{\sqrt{\pi}}  \left( \int_0^{\pi} + \int_{\pi}^{\infty}
    \right) \mathd tt^{- 1 / 2} \mathe^{- td}  \sum' \mathe^{- tR_i^2} \equiv
    E_{11} + E_{12}\\
    E_2 & = & - \frac{1}{A_c}  \int \mathd \mathbf{r} \frac{e^2}{\sqrt{r^2 +
    d^2}} = - \frac{1}{\sqrt{\pi} A_c}  \int_0^{\infty} \mathd t \int \mathd
    \mathbf{r} \mathe^{- tr^2} \mathe^{- td^2} t^{- 1 / 2} = -
    \frac{\sqrt{\pi}}{A_c}  \int_0^{\infty} \mathd tt^{- 3 / 2} \mathe^{-
    td^2}\\
    & = & - \frac{\sqrt{\pi}}{A_c}  \int_0^{\pi} \mathd tt^{- 3 / 2}
    \mathe^{- td^2} - \frac{2}{A_c}  \left( e^{- \pi d^2} - \pi d \text{erfc}
    \left( \sqrt{\pi} d \right) \right)
  \end{array}
\end{equation}
where $\Gamma (n) z^{- n} = \int_0^{\infty} t^{n - 1} \mathe^{- zt} \mathd t$
is used in rewriting $1 / \sqrt{d^2 + r^2} = \frac{1}{\sqrt{\pi}} 
\int_0^{\infty} t^{- 1 / 2} \mathe^{- t (d^2 + r^2)} \mathd t$, and
$\sqrt{\pi}  \int_{\pi}^{\infty} \mathd tt^{- 3 / 2} \mathe^{- td^2} =
\frac{2}{\sqrt{\pi}}  \left( e^{- \pi d^2} - \pi d \text{erfc} \left(
\sqrt{\pi} d \right) \right)$. $\sum'$ stands for the summation excluding $R_i
= 0$.

Let $t = \pi x$, we have
\begin{equation}
  E_{12} = \int_1^{\infty} \mathd xx^{- 1 / 2}  \sum\nolimits' \mathe^{- \pi x (d^2 +
  R_i^2)} = \sum\nolimits' \tmop{erfc} \left[ \sqrt{\pi (d^2 + R_i^2)} \right] /
  \sqrt{(d^2 + R_i^2)},
\end{equation}
where $\int_1^{\infty} x^{- 1 / 2} \mathe^{- \pi xa^2} \mathd x = \tmop{erfc}
\left( \sqrt{\pi} a \right) / a$ is utilized. To calculate $E_{11}$ we first
complete it with a $R_i = 0$ term
\begin{equation}
  \begin{array}{ccl}
    E_{11} & = & \frac{1}{\sqrt{\pi}}  \int_0^{\pi} \mathd tt^{- 1 / 2}
    \mathe^{- td^2} \Theta_{\Gamma} (t / \pi) - \frac{1}{\sqrt{\pi}} 
    \int_0^{\pi} t^{- 1 / 2} \mathe^{- td^2} \mathd t\\
    & = & \frac{\sqrt{\pi}}{A_c}  \int_0^{\pi} \mathd tt^{- 3 / 2} \mathe^{-
    td^2} \Theta_{\Gamma'} (\pi / t) - \tmop{erf} \left( \sqrt{\pi} d \right)
    / d\\
    & = & \frac{1}{A_c}  \int_1^{\infty} \mathd xx^{- 1 / 2} \mathe^{- \pi
    d^2 / x}  \sum' \mathe^{- \pi xK_i^2} - \tmop{erf} \left( \sqrt{\pi} d
    \right) / d + \frac{\sqrt{\pi}}{A_c}  \int_0^{\pi} \mathd tt^{- 3 / 2}
    \mathe^{- td^2}
  \end{array}
\end{equation}
with $\Theta_{\Gamma} (t) \equiv \sum_{\mathbf{R}_i \in \Gamma} \mathe^{- \pi
tR_i^2}, \Gamma$ being a lattice. From first line to second line we used
$\int_0^1 t^{- 1 / 2} \mathe^{- \pi td^2} \mathd t = \tmop{erf} \left( a
\sqrt{\pi} \right) / a$ and $\Theta_{\Gamma} (t) = t^{- n / 2} v (\Gamma)^{-
1} \Theta_{\Gamma'} (1 / t)$, where $\Gamma'$ is the dual of lattice $\Gamma$,
$v (\Gamma)$ is the measure of unit cell of $\Gamma$ and $n$ is dimension
of the lattice $\Gamma$ (see e.g. pg. 115 of \cite{Serre1973}); from
second line to third line, points of dual lattice are denoted as $\mathbf{K}_i$
and we let $t = \pi / x$ for all $K_i \equiv | \mathbf{K}_i | \neq 0$ terms.
Note that the very last divergent term in $E_{11}$ cancel the divergent part
of $E_2$.

Since
\begin{equation}
  \begin{array}{ccl}
    \int_1^{\infty} \mathd xx^{- 1 / 2} \mathe^{- \pi (d^2 / x + K_i^2 x)} & =
    & \frac{\mathe^{- 2 \pi dK_i}  \left( 1 + \tmop{erf} \left[ \sqrt{\pi}  (d
    - K_i) \right] \right) + \mathe^{2 \pi dK_i}  \left( 1 - \tmop{erf} \left[
    \sqrt{\pi}  (d + K_i) \right] \right)}{2 K_i}\\
    & \equiv & \phi_{- 1 / 2} (d, K_i)
  \end{array}
\end{equation}
we have
\begin{equation}
  \begin{array}{ccl}
    E_1 + E_2 & = & - \frac{\tmop{erf} \left( \sqrt{\pi} d \right)}{d} -
    \frac{2}{A_c}  \left( e^{- \pi d^2} - \pi d \text{erfc} \left( \sqrt{\pi}
    d \right) \right)\\
    &  & + \sum' \frac{\tmop{erfc} \left[ \sqrt{\pi (d^2 + R_i^2)}
    \right]}{\sqrt{d^2 + R_i^2}} + \frac{1}{A_c} \sum' \phi_{- 1 / 2} (d, K_i)
  \end{array}
  \label{eq:U1+U2}
\end{equation}
For a sanity check, let $d \rightarrow 0$ we have
\begin{equation}
  \begin{array}{ccl}
    E_1 + E_2 & = & - 2 \left( 1 + \frac{1}{A_c} \right) + \sum' \tmop{erfc}
    \left( \sqrt{\pi} R_i \right) / R_i + \frac{1}{A_c} \sum' \tmop{erfc}
    \left( \sqrt{\pi} K_i \right) / K_i\\
    & \cong & - 2 \left( 1 + \frac{1}{A_c} \right) + 6 \tmop{erfc} \left(
    \sqrt{\pi} \right) + 6 \tmop{erfc} \left( \sqrt{\pi} / A_c \right)\\
    & = & - 4.213475
  \end{array}
\end{equation}
where we took nearest lattice point approximation, i.e. only six terms with smallest
$R_i,K_i$ in the those lattice summations are kept. Nevertheless the result match
the known static energy for 2d Wigner crystal up to fourth digit.

For general $d$, let $\delta E (d)$ be the nearest lattice point approximation of $E_1 + E_2$
in \Cref{eq:U1+U2}
\begin{equation}
  \begin{array}{ccl}
    \delta E (d) & = & - \frac{\tmop{erf} \left( \sqrt{\pi} d \right)}{d} - \frac{2
    \left( e^{- \pi d^2} - \pi d \text{erfc} \left( \sqrt{\pi} d \right)
    \right)}{A_c} + \frac{6 \tmop{erfc} \left[ \sqrt{\pi (d^2 + 1)}
    \right]}{\sqrt{d^2 + 1}}\\
    &  & + 3 \left\{ \mathe^{- 2 \pi d / A_c}  \left( 1 + \tmop{erf} \left[
    \sqrt{\pi}  \left( d - \frac{1}{A_c} \right) \right] \right) + \mathe^{2
    \pi d / A_c} \tmop{erfc} \left[ \sqrt{\pi}  \left( d + \frac{1}{A_c}
    \right) \right] \right\}
  \end{array}
  \label{eq:deltaU}
\end{equation}
$1 / A_c = 2 / \sqrt{3}$ comes from lattice constant of the dual lattice. It behaves
asymptotically in the $d \rightarrow \infty$ limit as $\delta E (d) + 1 / d \sim 6
\mathe^{- 4 \pi d / \sqrt{3}}$. Also for $d \rightarrow \infty$, $\tmop{erfc}
\left[ \sqrt{\pi (d^2 + R_i^2)} \right] / \sqrt{d^2 + R_i^2} \sim \mathe^{-
\pi (d^2 + R_i^2)} / (\pi (d^2 + R_i^2))$ and $\tmop{erfc} \left( \sqrt{\pi} x
\right) \sim \mathe^{- \pi x^2} / (\pi x)$ results in
\begin{equation}
  \phi_{- 1 / 2} (d, K_i) \leqslant [2 \mathe^{- 2 \pi dK_i} + \mathe^{- \pi
  (d^2 + K_i^2)} / (\pi (d + K_i))] / (2 K_i),
\end{equation}
All terms generated from farther lattice points are dominated by $6 \mathe^{-
4 \pi d / \sqrt{3}}$. In the sense that $\delta E (d)$ is a good approximation to
$E_1 + E_2$ for both $d \rightarrow 0$ and $d \rightarrow \infty$, we could
safely take
\begin{equation}
  \Delta E_e \cong 1 / d + \delta E (d)
\end{equation}

Putting back dimensions, the Coulomb energy density difference is, with $\delta E$ defined in \Cref{eq:deltaU}, 
\begin{equation}
    \Delta \varepsilon_e \cong n_e \frac{Q^2}{a} (a / d + \delta E (d/a)) \label{e-coulomb}
\end{equation}

\subsection{Phase Diagram}

\begin{figure}[!h]
    \resizebox{.95\linewidth}{!}{
        \includegraphics{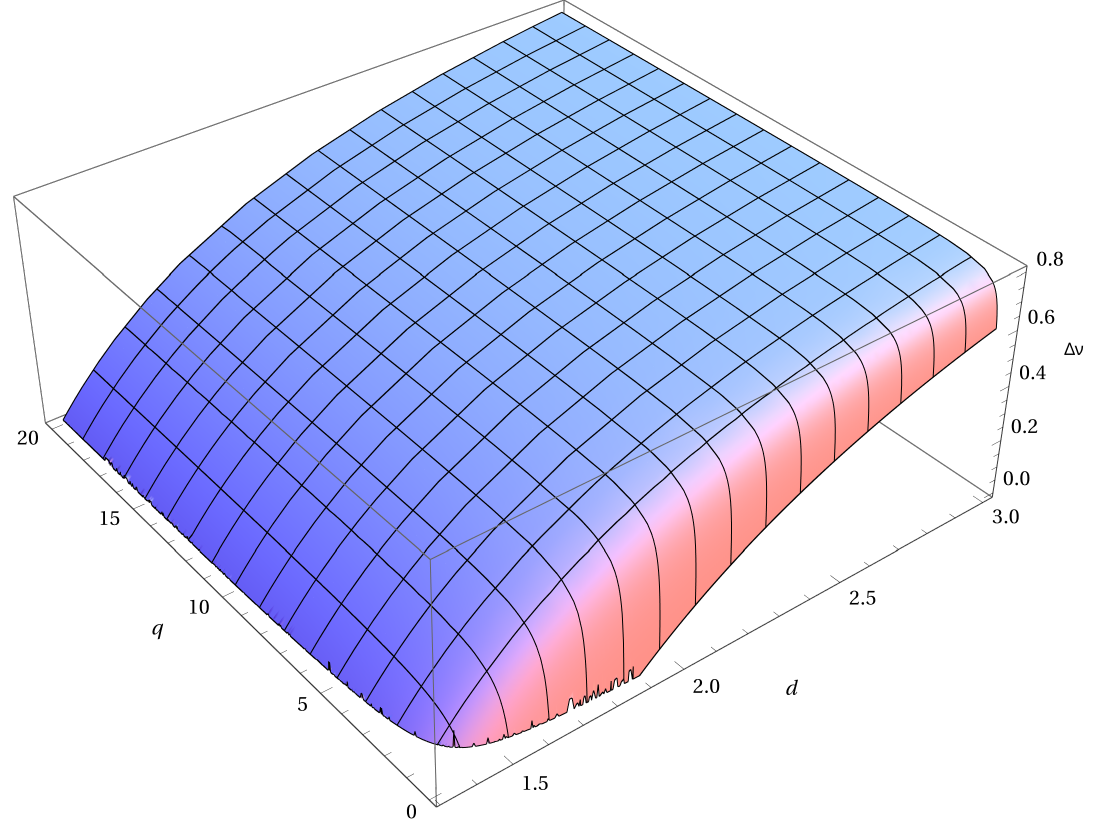} 
    }
    \caption{Phase diagram of coupled/decoupled Wigner crystal plotted from \Cref{eq:trans2}.
    $q=\frac{4 \Delta^2}{\sqrt{3} Q^4}$ characterizes the random potential strength,
    where $\Delta$ is defined in \Cref{eq:random-potential} and
    $Q^2=e^2/(4\pi\epsilon)$. Region under the surface is decoupled electron-hole Wigner 
    crystal while region above it is exciton Wigner crystal. \label{fig:p-d-2}}
\end{figure}

Comparing \eqref{eq:e-r} with \eqref{e-coulomb} we can immediately see that
the transition between coupled/decoupled phases is determined by the root of
the dimensionless equation
\begin{equation}
    \frac{q}{0.3} x^2 - x - q/2.5 - x^2 \delta E(x)=0, 
    \quad x=d/a = d \sqrt{\frac{\sqrt{3}}{8\pi} (1-\Delta \nu)} \label{eq:trans2}
\end{equation}
where $q$, defined in \Cref{eq:def:q}, is, up to a constant,
the energy scale of random potential comparing with
Coulomb energy. Putting together, we can draw a phase diagram 
\Cref{fig:p-d-2} for the decoupled electron-hole Wigner crystal and exciton Wigner crystal.




\section{Concluding Remarks\label{summary}}

In this paper, we analyzed the competition between different phases in a bilayer
quantum hall system with total filling factor 1 driven by temperature and/or
disorder. Our results compare favorably with a recent experiment
\cite{zeng2023}. Particularly interesting (and surprising) among our findings is
that the exciton superfluid can (often) result from melting an exciton WC. This
bears remarkable similarity to the observation \cite{Pan_2002} that melting of
electron WC at low filling factor results in fractional quantum Hall
liquids. Similar phenomena was observed very recently in systems supporting
(fractional) anomalous quantum Hall states \cite{Lu_2024}. We speculate that
melting of electron or hole WC in these systems resulted in the formation of
fractional anomalous quantum Hall states. We also note that it is in principle
possible to have the WC and SF orders coexist, resulting in an exciton
supersolid. It is a very interesting future direction of research to look for
such a novel phase, both experimentally and theoretically.

\section*{ACKNOWLEDGMENTS}

We thank Cory Dean, Lloyd Engel and Leo Li for helpful discussions. This work was supported by the National Science Foundation Grant No. DMR-2315954, and performed at the National High Magnetic Field Laboratory which is supported by National Science Foundation Cooperative Agreement No. DMR-2128556, and the State of Florida.

\bibliography{report}

\end{document}